\documentclass[useAMS,usenatbib,referee]{mn2e}

\usepackage{graphicx}
\usepackage{amsmath}
\usepackage{bm}
\usepackage{epsfig}
\usepackage{amssymb}
\usepackage{natbib}
\usepackage{threeparttable}
\usepackage{subfigure}
\usepackage{setspace}
\usepackage{booktabs,multirow}
\usepackage{appendix}
\usepackage{rotating}
\usepackage{lscape}

\def\simmore{\mathbin{\lower 3pt\hbox
     {$\rlap{\raise 5pt\hbox{$\char'076$}}\mathchar"7218$}}}   

\title[Investigating mHz QPOs in 4U 1636$-$53]{Spectral and timing analysis of the mHz QPOs in the neutron-star low-mass X-ray binary 4U 1636--53}

\author[Ming Lyu et al.]
{Ming Lyu$^1$\thanks{E-mail: m.lyu@astro.rug.nl}, Mariano M\'endez$^1$, Guobao Zhang$^2$, L.~Keek$^3$ \\
$^1$Kapteyn Astronomical Institute, University of Groningen, PO BOX 800, NL-9700 AV Groningen, the Netherlands\\
$^2$New York University Abu Dhabi, P.O. Box 129188, Abu Dhabi, United Arab Emirates\\
$^3$Center for Relativistic Astrophysics, School of Physics, Georgia Institute of Technology, Atlanta, GA 30332, USA
}

\begin{document}

\date{Accepted XXXX. Received XXXX; in original form XXXX}

\maketitle

\label{firstpage}

\begin{abstract}
We investigate the spectral and timing properties of the millihertz quasi-periodic oscillations (mHz QPOs) in neutron-star low-mass X-ray binary 4U 1636--53 using XMM-Newton and Rossi X-ray Timing Explorer (RXTE) observations. The mHz QPOs in the XMM-Newton/RXTE observations show significant frequency variation and disappear right before type I X-ray bursts. We find no significant correlation between the mHz QPO frequency and the temperature of the neutron-star surface, which is different from theoretical predictions. For the first time we observed the full lifetime of a mHz QPO lasting 19 ks. Besides, we also measure a frequency drift timescale $\sim$15 ks, we speculate that this is the cooling timescale of a layer deeper than the burning depth, possibly heated by the previous burst. Moreover, the analysis of all X-ray bursts in this source shows that all type I X-ray bursts associated with the mHz QPOs are short, bright and energetic, suggesting a potential connection between mHz QPOs and He-rich X-ray bursts. \\

\end{abstract}


\section{introduction}
The millihertz quasi-periodic oscillations (mHz QPOs) in neutron-star low-mass X-ray binaries (LMXBs) were first detected in 4U 1608--52, 4U 1636--53, and Aql X-1 by \citet{revni01}. The mHz QPOs occur only within a narrow range of X-ray luminosities, $L_{\rm 2-20 keV} \simeq (5-11) \times 10^{36}$ ergs s$^{-1}$, in the frequency range $7-9$ mHz with strong flux variation below 5 keV. Besides, the mHz QPOs often disappear right before a type I X-ray burst \citep{revni01,diego08}.

It has been proposed that the mechanism responsible for the mHz QPOs is connected with some sort of nuclear burning on the neutron star surface. For instance, \citet{revni01} suggested that the mHz QPOs resulted from a special mode of nuclear burning on the neutron star surface when the source was within a certain range of mass accretion rate. \citet{heger07} further proposed that the mHz QPOs were due to marginally stable nuclear burning of Helium on the surface of accreting neutron stars, with a predicted oscillation timescale of $\sim$100 s, consistent with the observed $\sim$2 minutes period of the mHz QPOs.

Several results have strengthened the nuclear burning explanation of the mHz QPOs. \citet{yu02} detected mHz QPOs and kilohertz QPOs in the same observations of 4U 1608--52. The $2-5$ keV X-ray count rate in the mHz QPO cycles was anti-correlated with the frequency of the kilohertz QPOs. This result was consistent with the nuclear burning origin of the mHz QPOs: As the luminosity increases, the stresses of radiation coming from the neutron star surface push the inner disc outward so that the frequency of the kilohertz QPOs decreases. \citet{diego08} found a systematic frequency drift of the mHz QPO until the QPO disappeared before a type I X-ray burst in 4U 1636$-$53, indicating a connection between the mHz QPOs and the unstable nuclear reaction on the neutron star surface.

 \citet{linares10} reported a new detection of mHz QPOs at $\sim$4.5 mHz in the neutron star transient IGR J17480-2446 at a persistent luminosity $L_{2-50 keV}$ $\sim 10^{38}$ erg s$^{-1}$, about one order of magnitude higher than observed in the previous mHz QPO sources. In addition, \citet{linares12} found a transition between thermonuclear bursts and mHz QPOs in this source: the X-ray bursts gradually evolved into the mHz QPO as accretion rate increased, and vice versa. This behaviour was similar to the prediction from the one-zone models and simulation results of the marginally stable burning of \citet{heger07}, further confirming that the mHz QPO was connected with marginally stable burning on the neutron star surface. 

In the model of \citet{heger07}, the required mass accretion rate for marginally stable burning is $\sim$ $\dot{\rm M}_{\rm Edd}$, about ten times higher than the rate deduced by the average X-ray luminosity \citep{revni01,yu02,diego08}. \citet{keek09} found that the critical accretion rate at which mHz QPOs were observed could be explained if they considered the turbulent chemical mixing of the accreted fuel together with a higher heat flux from the crust. Furthermore, the cooling of the deep layers also explains the observed frequency drifts before X-ray bursts. Recently, \citet{keek14} investigated the influence of the chemical composition of the accreted fuel and nuclear reaction rates on the marginally stable burning process; their results showed that no allowed variation in accretion fuel composition and nuclear reaction rate was able to produce marginally stable burning at the observed accretion rate.

\citet{lyu14b} found that the frequency of the mHz QPOs correlates with the temperature of the neutron star surface in 4U 1636--53 using an XMM-Newton observation. This frequency-temperature correlation agrees well with the results of \citet{keek09}: As the burning layers cool down, the frequency of the oscillations decreases by tens of per cents until a burst happens. Besides, \citet{lyu14b} found that the QPO in that observation disappeared after one burst, and it subsequently reappeared $\sim$25.3 ks later, which was the longest reappearance time scale so far measured for a mHz QPO in any LMXBs.

In this work we investigate the frequency-temperature correlation in the mHz QPO cycle using XMM-Newton observations of the LMXB 4U 1636--53 to take a closer look into the mechanism behind the mHz QPOs. Using simultaneous observations with the Rossi X-ray Timing Explorer (RXTE), this is the first time that we are able to carry out a wide energy band (0.8$-$100 keV) spectral analysis in the mHz QPO cycles. Furthermore, we explored the possible relation between the mHz QPOs and the associated X-ray bursts, which may provide a new way to investigate the origin of mHz QPOs. In Section 2 we describe the observations and data analysis, and in Section 3 we present our main results. Finally, we discuss the results in the framework of the marginally stable burning model in Section 4.

\section{observations and data reduction}
  The three XMM-Newton observations of 4U 1636--53 that we use in this work were performed with the European Photon Imaging Camera, EPIC-PN \citep{xmm01} in the timing mode, with a time resolution of 0.03 ms and in the energy band 0.5$-$12 keV. The three observations were taken between 2007 to 2009, with a total duration of about 90 ks. We selected RXTE observations that were taken simultaneously with those three XMM-Newton observations. The Proportional Counter Array \citep[PCA;][]{jahoda06} and the High Energy X-ray Timing Experiment \citep[HEXTE;][]{roth98} onboard RXTE covered the energy range 2$-$60 keV and 15$-$250 keV, respectively. Both the PCA and HEXTE data were included in this work in order to get a wide energy band coverage. We also re-analyzed the XMM-Newton observation (ObsID 0606070301) in \citet{lyu14b} with the latest calibration files, and selected the simultaneous RXTE observation (ObsID 94310-01-03-000) for the spectral analysis. 
    
 In this work, we used the Science Analysis System (SAS) version 14.0.0 for the XMM-Newton data reduction, with the latest calibration files applied (as of April 2015). We selected the Rate-Dependent PHA (RDPHA) method in the command {\tt epproc} for the correction of energy scale rate dependence in EPIC-PN Timing Mode exposures. We applied the task {\tt barycen} to convert the arrival time of photons from the local satellite frame to the barycenter of the Solar system. We found that there was some moderate pile up in all XMM-Newton observations according to the results of the {\tt epatplot} test. We finally selected a 10-column wide region centered at the position of the source, excluding the central column for Obs 1 and the three central columns for the other observations. We only selected single and double events ({\sc pattern} $\le$ 4) to extract spectra and light curves.

Furthermore, we investigated all X-ray bursts of 4U 1636--53 detected by the PCA \citep{jahoda06} in order to explore the possible connection between mHz QPOs and the associated X-ray bursts. For this we produced 0.25-s light curves from the Standard-1 data and searched for X-ray bursts in these 
light curves following the procedure described in \cite{zhang11}. A total of 338 type-I X-ray  
bursts have been detected in these data. 

We used the Standard-2 data to calculate the source X-ray colours. We defined
soft and hard X-ray colours as the $3.5-6.0/2.0-3.5$ keV and $9.7-16.0/6.0-9.7$
keV count rate ratios, respectively \citep[see][for details]{zhang11}. We
parametrized the position of  the source on the colour-colour diagram (CCD) by the length
of a solid curve $S_{\rm a}$ \citep[see, e.g. ][]{Mendez99, zhang11}, fixing the values
of $S_{\rm a}=1$ and $S_{\rm a}=2$ at the top-right and the bottom-left vertex of the
CCD, respectively.

\begin{figure}
\center
\includegraphics[height=0.5\textwidth]{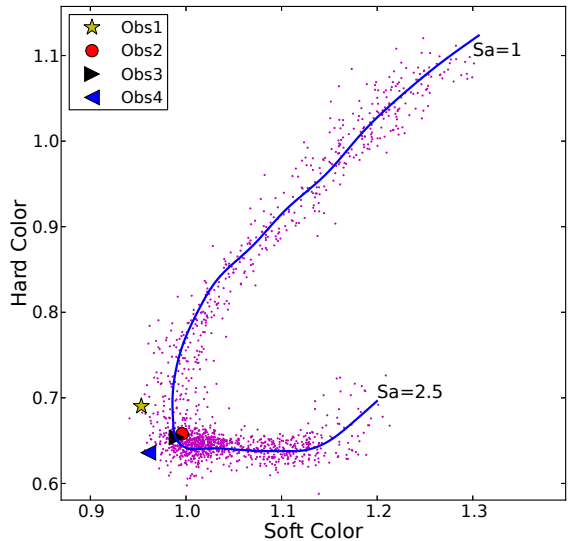}
\caption{Colour-colour diagram of 4U 1636--53. Each magenta point (grey in the black and white version of this figure) represents the averaged Crab-normalised colours \citep[see][for details]{zhang11} of a single RXTE observation. In the legend we indicate the observations used in this work. The position of the source in the diagram is parameterised by the length of the blue solid curve $S_{\rm a}$, the details of $S_{\rm a}$ curve is shown in the $\S$2.}
\label{ccd}
\end{figure}

The colour-colour diagram in Figure \ref{ccd} shows that the four XMM-Newton observations were performed when the source was in the transitional state between the hard and the soft spectral state. The details of the XMM-Newton and simultaneous RXTE observations in this work are listed in Table \ref{overall}.

\begin{table*}
\caption{XMM-Newton / simutaneous RXTE observations of 4U 1636$-$53 used in this work.}
\begin{tabular}{|c|c|c|c|c|}
\hline
\hline
Observation   &   Instrument  & Observation ID    & Start Time   &    Observation Length (ks)    \\
\hline
 Obs 1       &   XMM-Newton   &   0500350301      & 2007-09-28 15:43 &   30.6 \\
             &    RXTE        &   93091-01-01-000 & 2007-09-28 14:48 &   28.8 \\
             &    RXTE        &   93091-01-01-00  & 2007-09-28 22:48 &   9.5 \\
             &                &                   &                  &    \\
 Obs 2       &   XMM-Newton   &   0500350401      & 2008-02-27 04:14 &   38.6 \\
             &   RXTE         &   93091-01-02-000 & 2008-02-27 03:47 &   28.8 \\
             &   RXTE         &   93091-01-02-00  & 2008-02-27 11:47 &   10.0 \\
             &                &                   &                  &    \\
 Obs 3       &   XMM-Newton   &   0606070101      & 2009-03-14 15:11 &   23.8\\
             &   RXTE         &   94310-01-01-02  & 2009-03-14 15:05 &   2.4 \\
             &   RXTE         &   94310-01-01-04  & 2009-03-14 16:40 &   4.0 \\
             &   RXTE         &   94310-01-01-00  & 2009-03-14 18:18 &   13.4 \\
             &                &                   &                  &    \\
 Obs 4$^{*}$   &   XMM-Newton   &   0606070301      & 2009-09-05 01:57 &   40.9  \\
                       &   RXTE         &   94310-01-03-000 & 2009-09-05 01:17 &   23.2\\

\hline
\end{tabular}
\medskip
\\
$\ast$ We re-analysed the spectral properties of this observation in this work. For the timing properties of this observation, please refer to \citet{lyu14b}. \\
\label{overall}
\end{table*}

\subsection{Timing data}
\begin{figure}
\center
\includegraphics[height=0.85\textwidth]{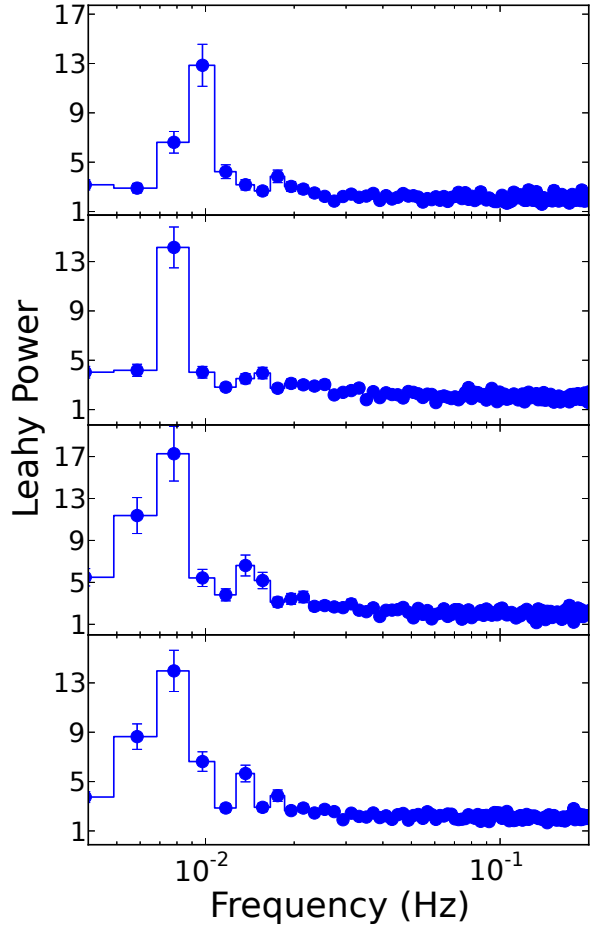}
\caption{Average power spectra of the four XMM-Newton observations of 4U 1636--53 (Obs 1 to 4 from top to bottom) in the $0.2-5$ keV range, calculated from the PN light curve at 1-second resolution. A significant QPO at about 8$-$10 mHz and a weak second harmonic are apparent in each observation.}
\label{ps}
\end{figure}

\begin{figure}
\center
\includegraphics[height=0.37\textwidth]{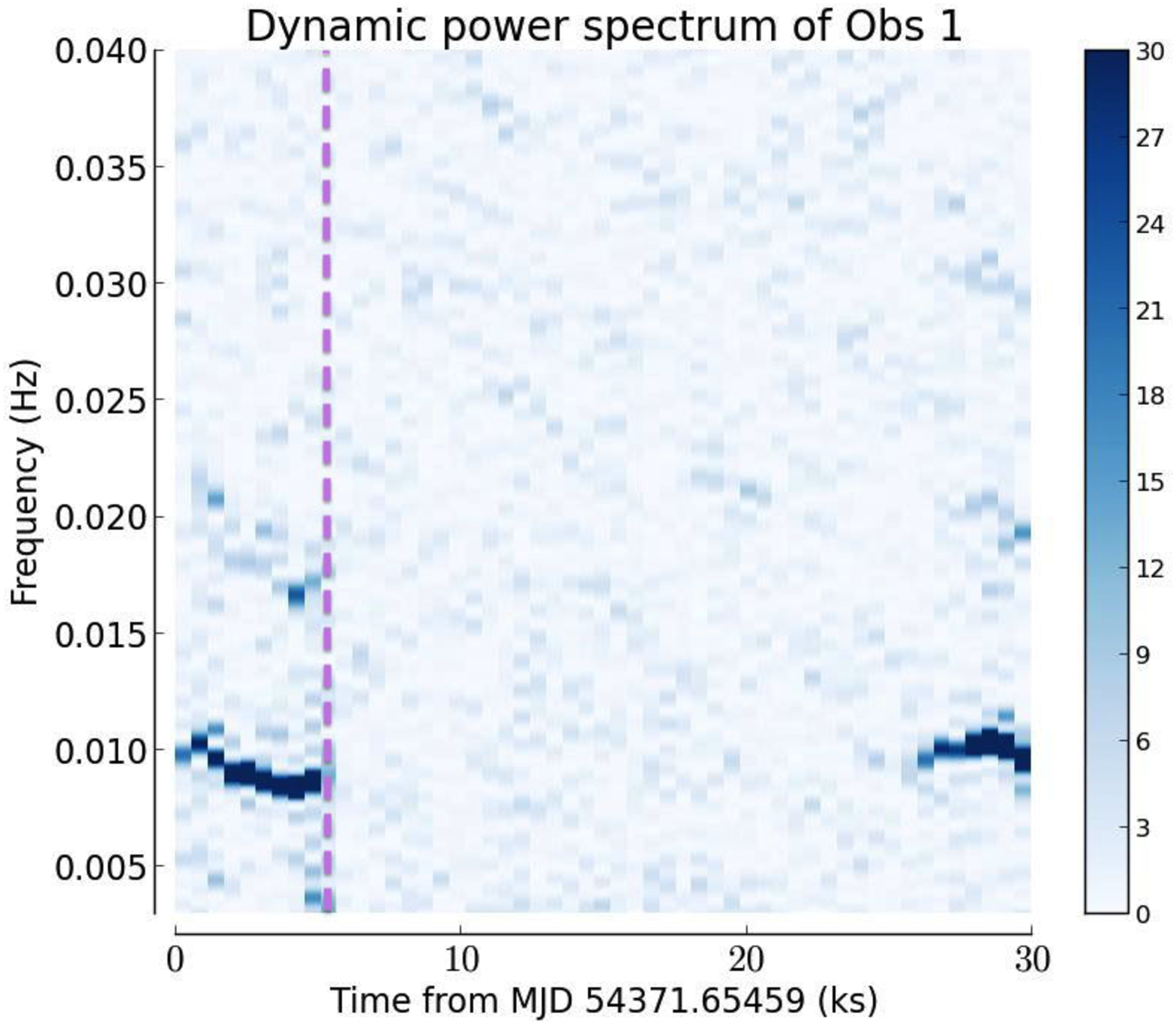}
\includegraphics[height=0.37\textwidth]{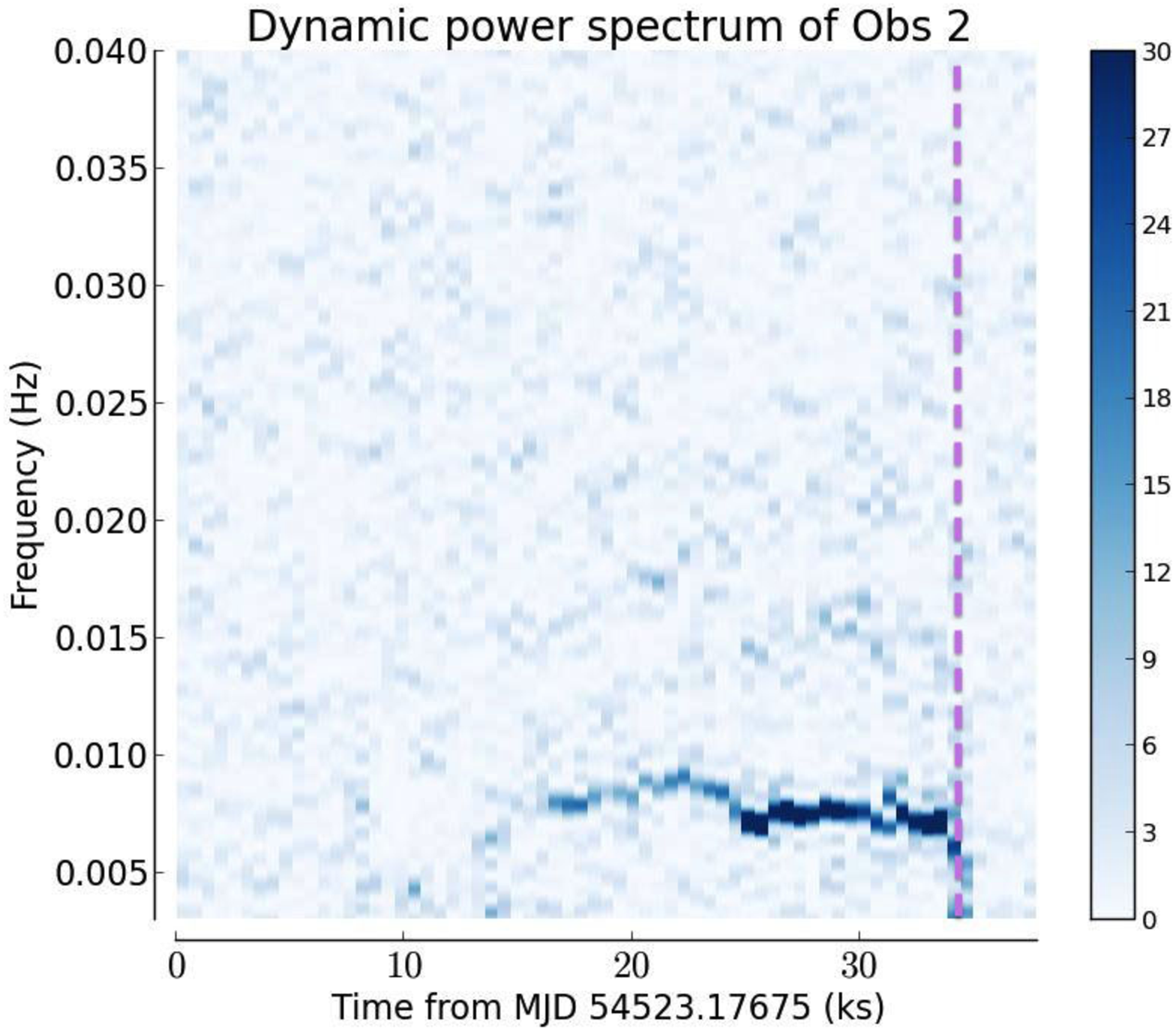}
\includegraphics[height=0.37\textwidth]{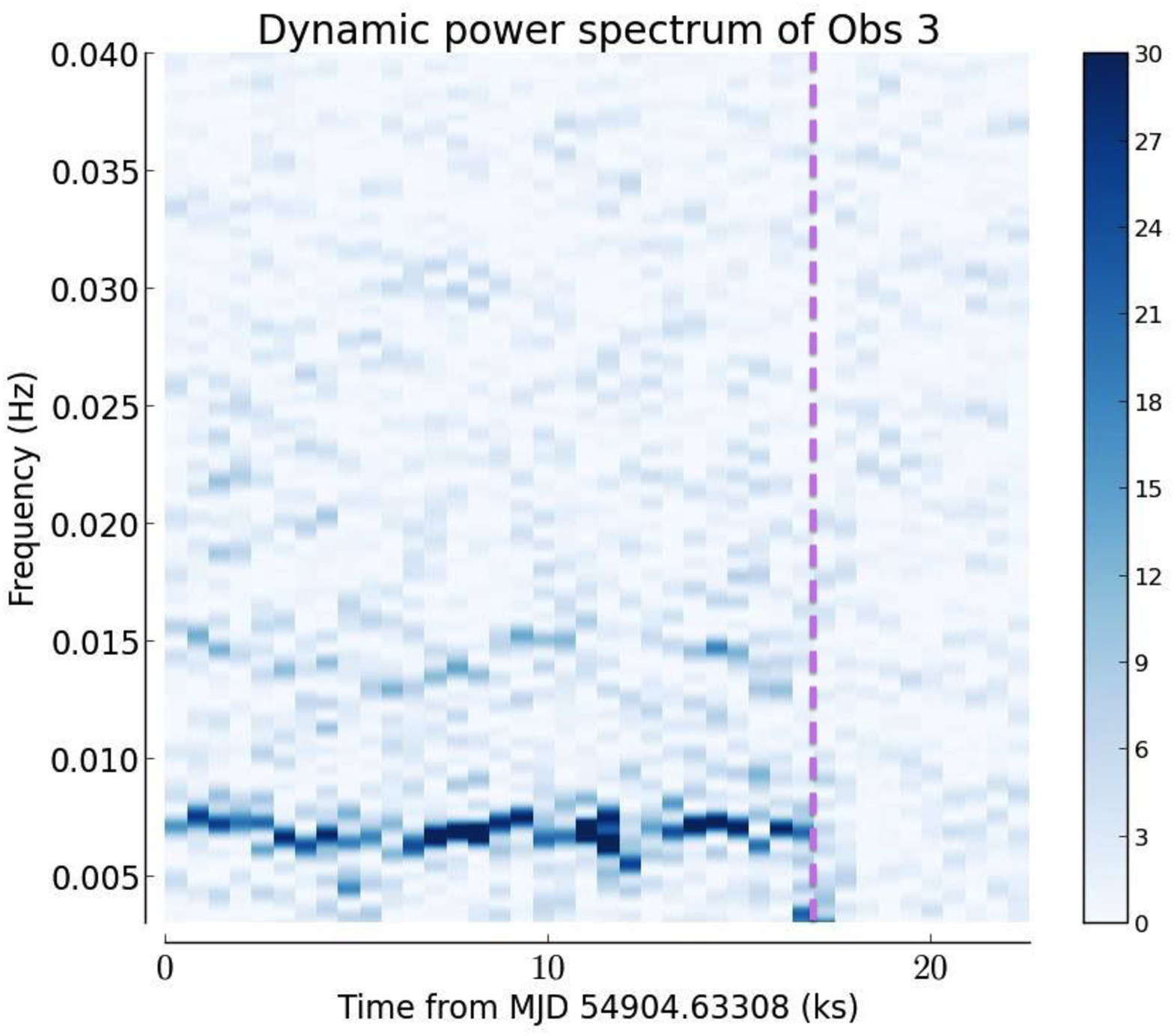}
\includegraphics[height=0.38\textwidth]{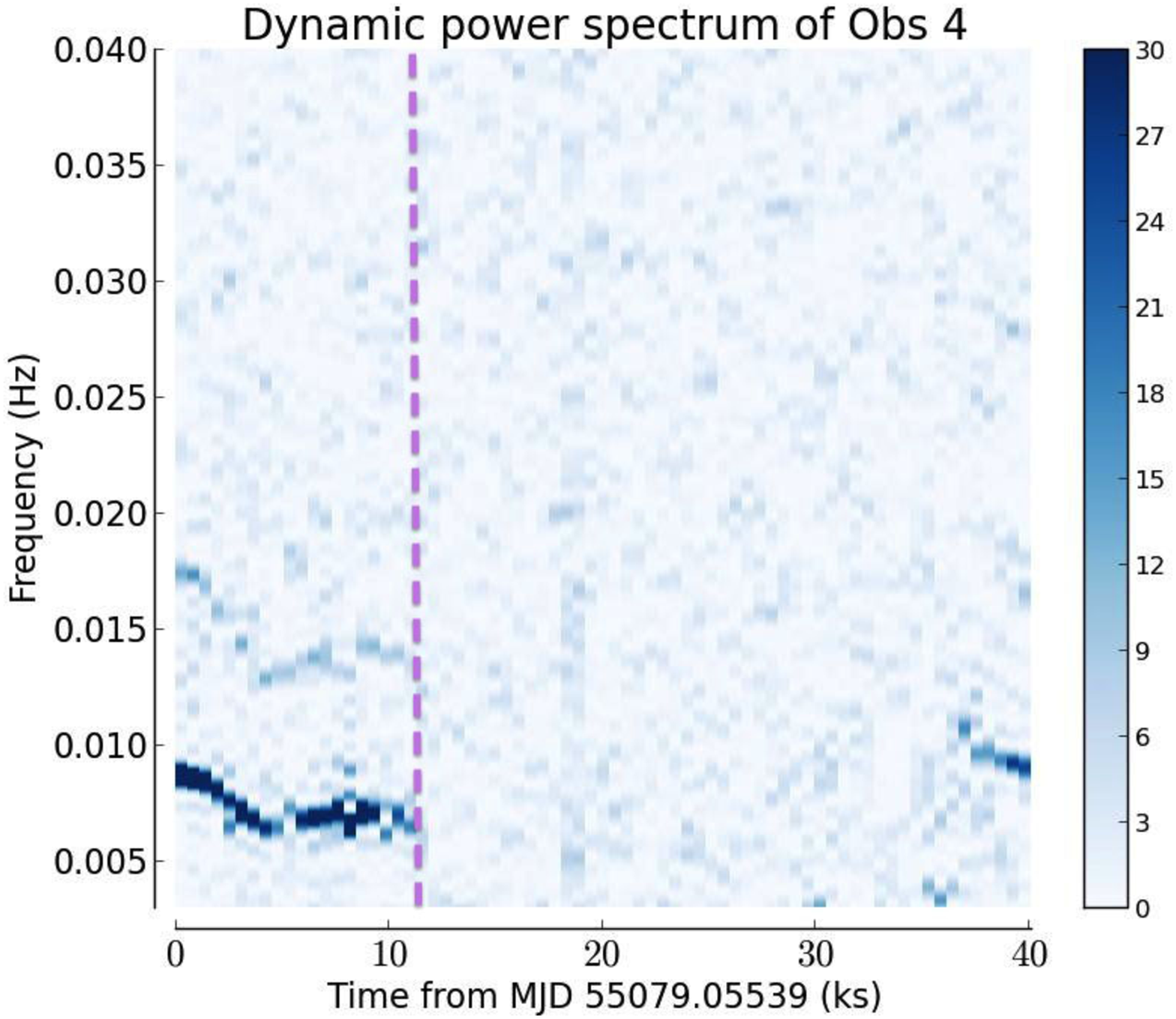}
\caption{Dynamic power spectra of the XMM-Newton observations of 4U 1636--53. The purple dashed lines indicate the time of X-ray bursts in these observations. Each column represents the power spectrum extracted from a time interval of 1130 seconds with the starting time of each interval set to 565 s after the starting time of the previous one. We oversampled the frequency scale by a factor of 100 using the Lomb-Scargle periodogram to display the evolution of the mHz QPOs frequency. We fixed the count rate during X-ray bursts and instrument dropouts to the average count rate. The colour bars on the right indicate the power at each frequency as defined in the Lomb-Scargle periodogram.}
\label{dps}
\end{figure}

We extracted a 1-s resolution light curve\footnote{We did not exclude the central columns of the CCD to produce these light curves in order to retain enough source photons for the detection of the mHz QPO in the power and dynamic power spectra.} in the 0.2$-$5 keV band and produced an average power spectrum (Figure \ref{ps}) for each XMM-Newton observation, after excluding instrument dropouts and X-ray bursts. We used the ISIS Version 1.6.2-27 \citep{houck2000} to compute the Fourier transform to intervals of 512-s duration using the command {\tt sitar$\_$avg$\_$psd}, in a frequency range of 1.95$\times$10$^{-3}$ to 0.5 Hz. To explore the frequency evolution of the mHz QPO, we generated a dynamic power spectrum with the frequency scale being oversampled by a factor of 100 using the Lomb-Scargle periodogram method \citep{lomb76,scargle82}. The dynamic power spectra of the four XMM-Newton observations are shown in Figure \ref{dps}; each column in those plots represents the power spectrum extracted from a time interval of 1130 seconds with the starting time of each interval set to 565 s after the starting time of the previous one. 

\begin{figure}
\center
\includegraphics[height=0.35\textwidth]{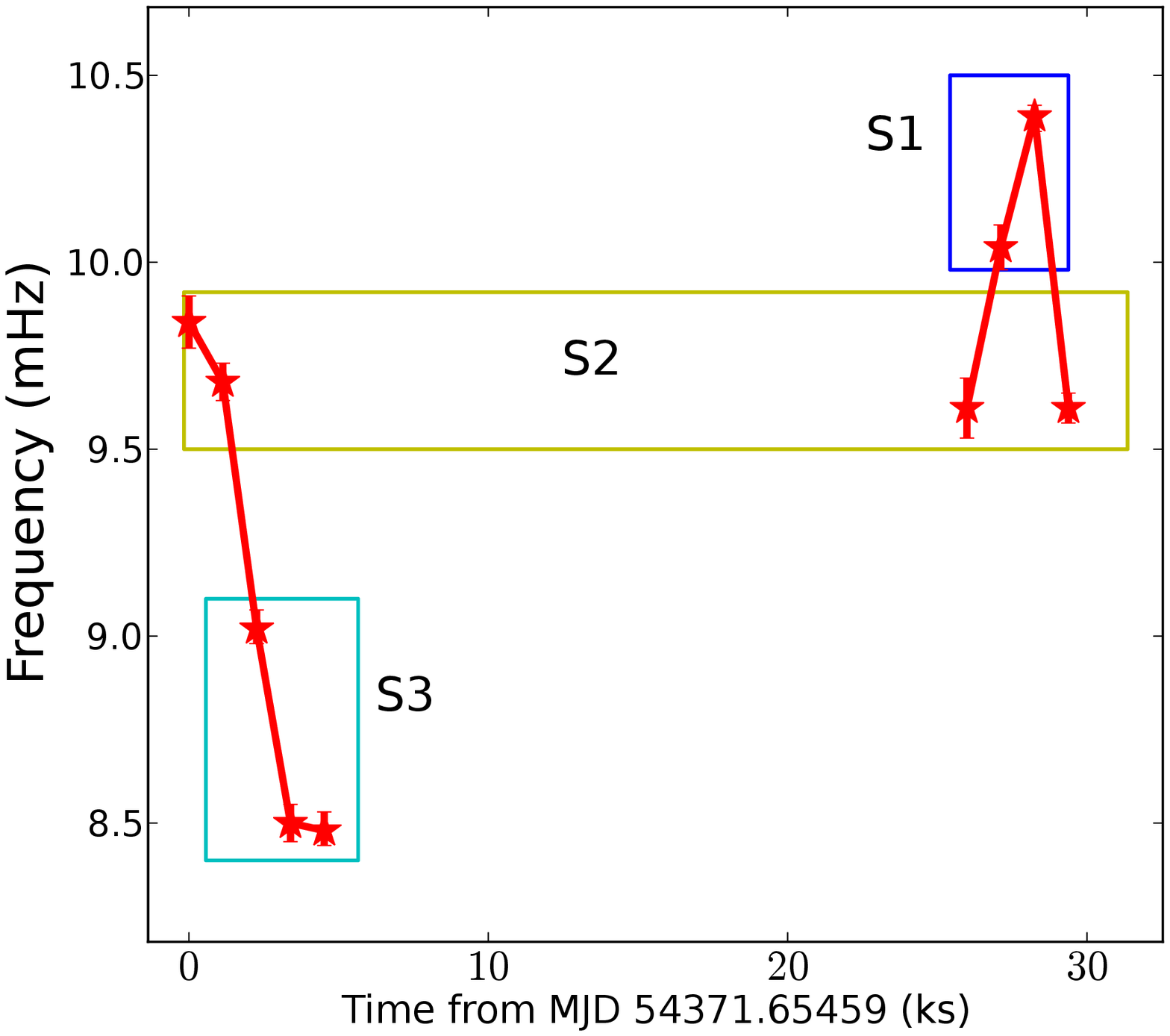}
\includegraphics[height=0.35\textwidth]{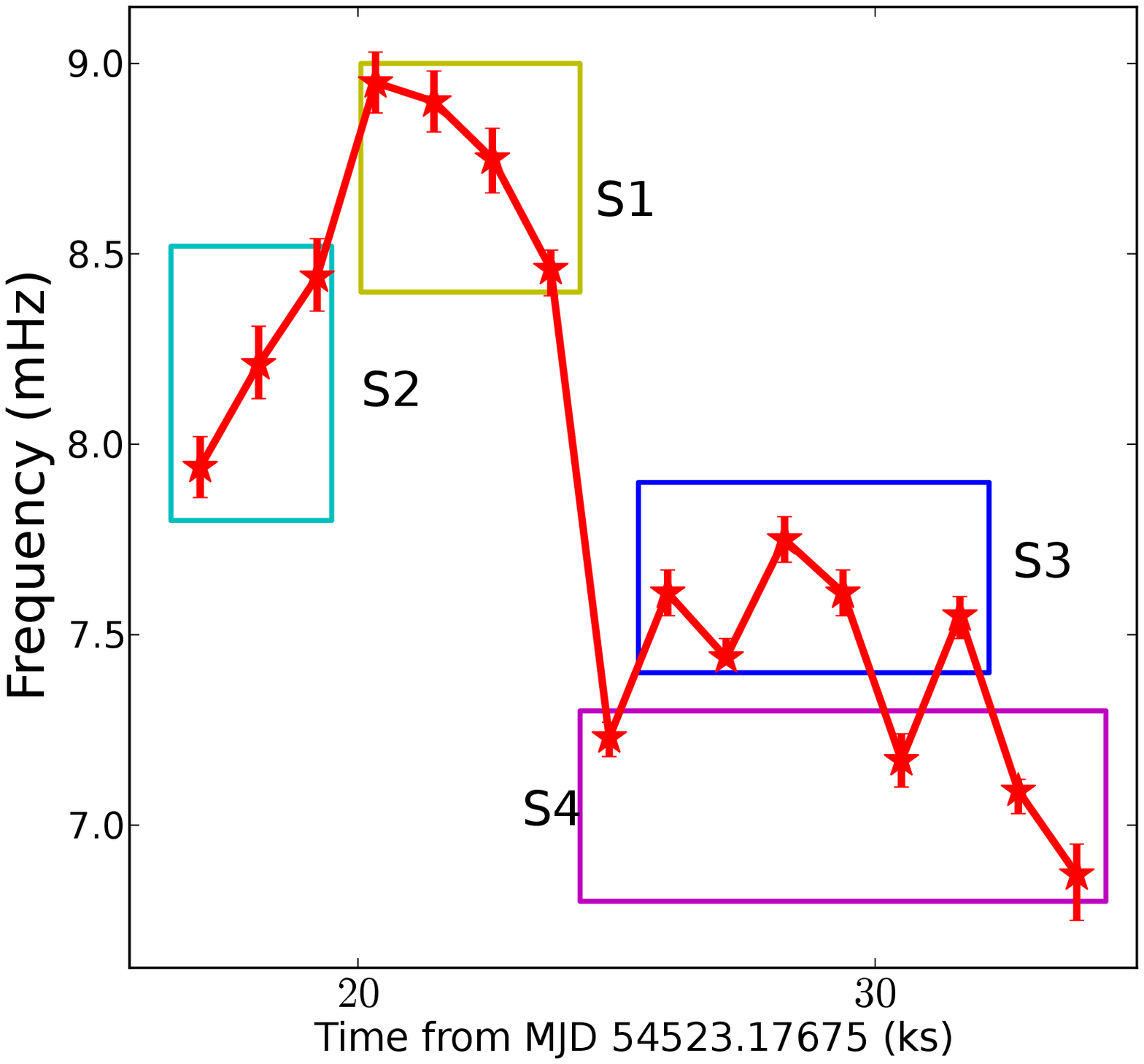}
\includegraphics[height=0.35\textwidth]{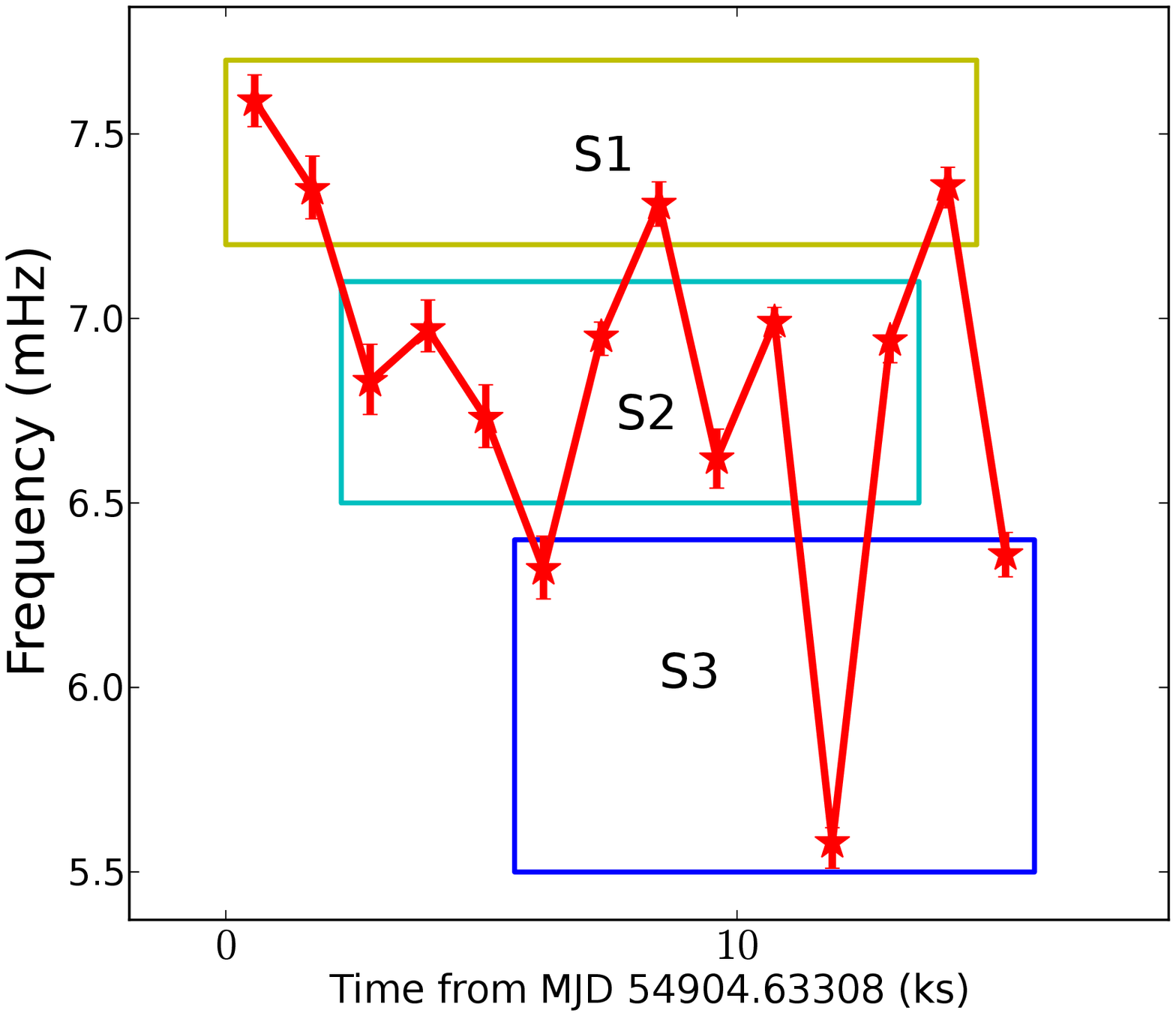}
\includegraphics[height=0.35\textwidth]{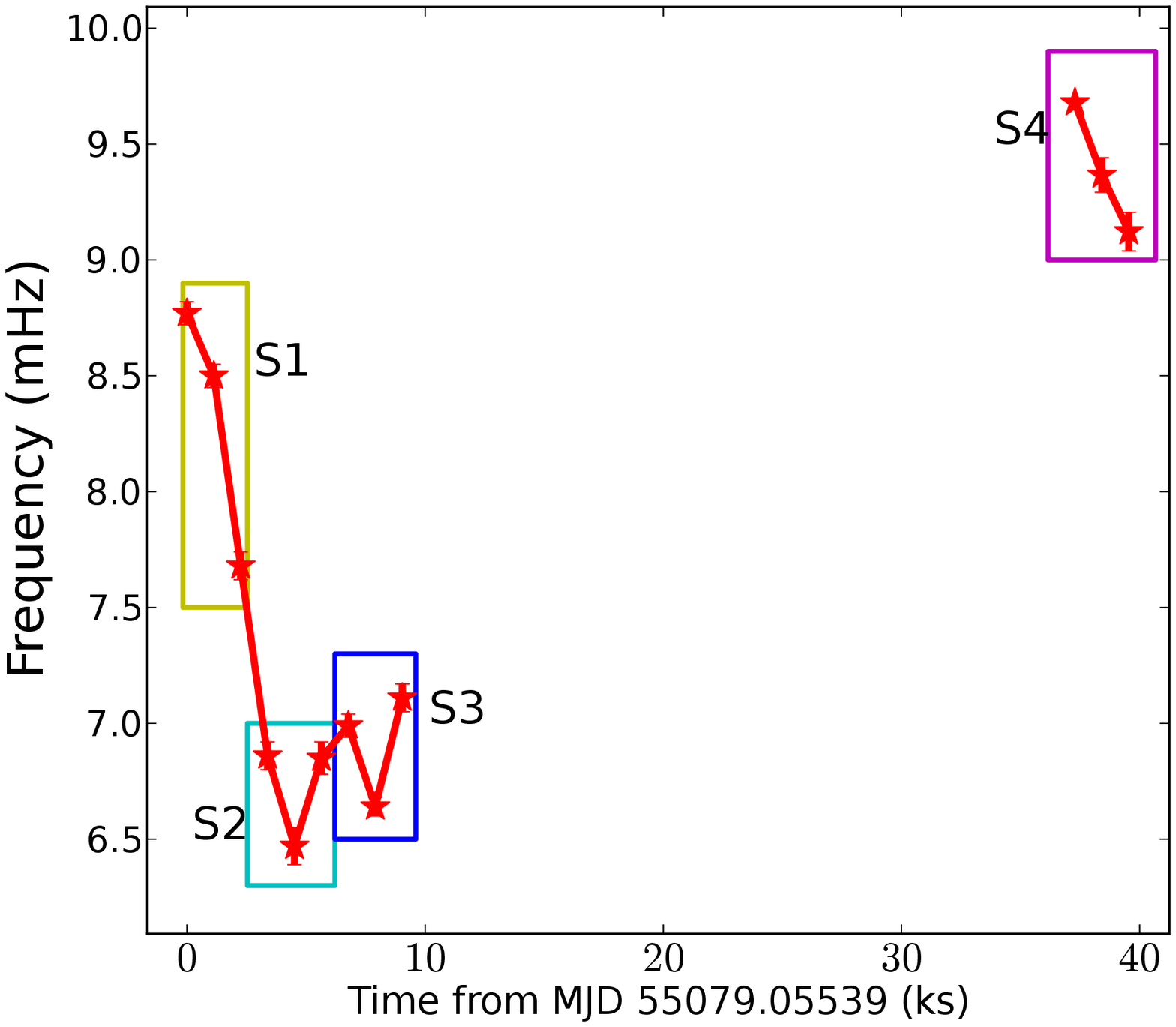}

\caption{Frequency evolution of the mHz QPO in the XMM-Newton observations of 4U 1636--53 (Obs 1 to 2 from left to right at the top; Obs 3 to 4 from left to right at the bottom). Each symbol corresponds to the frequency derived from the fit of each 1130 s independent light curve using a constant plus a sine function. We divided the QPO cycles in each observation into several segments according to the frequency of the QPO. The time intervals within which the QPO showed similar frequency were assigned the same segment number as indicated by the rectangles with segment number (S1-4) in each observation. Since the lengths of the segments in Obs 4 are not multiple of 1130 s (see Table \ref{segs} for details), here we round off the time span of each segment in Obs 4 to be a multiple of 1130 s for clarity.}
\label{fre}
\end{figure}

We fitted the light curves\footnote{We did exclude the central columns of the CCD to produce these light curves in order to measure the frequency and rms amplitude of the QPO.} of each independent 1130 s time interval in the XMM-Newton observations with a constant plus a sine function to map the frequency evolution of the mHz QPO accurately. The frequency behaviour of the mHz QPO in the XMM-Newton observations is shown in Figure \ref{fre}. Furthermore, in order to explore the possible mechanism responsible for the variation of the QPO frequency, for each XMM-Newton observation we divided the intervals with mHz QPO into several segments according to the frequency of the QPO (Figure \ref{fre}), and calculated the average frequency and average rms amplitude of the QPO in each segment; the details of the segments are listed in Table \ref{segs}. For the timing reduction and analysis of the Obs 4, please refer to \citet{lyu14b}.

\begin{table*}
\caption{Information for the segments of XMM-Newton observations of 4U 1636$-$53. Here we give the standard deviation of the average frequency since the frequency in segments may change a lot. All errors in the Tables are at the 90\% confidence level unless otherwise indicated. We show the information of the segments in the Obs 4 from \citet{lyu14b}.}
\begin{tabular}{|c|c|c|c|c|}
\hline
\hline
Observation   &   Segment number & Intervals (s)  &    Average frequency (mHz)  & rms amplitude (\%)   \\
\hline
 Obs 1       &    S1   & 27120-29380                                                &  10.2$\pm$0.3     & 1.4$\pm$0.2\\
                  &    S2   &  0-2260;  25990-27120; 29380-30510         &  9.7$\pm$0.1       &  1.2$\pm$0.1        \\
                  &    S3   & 2260-5650                                                    &   8.7$\pm$0.3      &  1.6$\pm$0.2\\
 \\
 Obs 2       &    S1   &    20340-24860                                             &  8.7$\pm$0.2       & 0.8$\pm$0.2\\
                  &    S2   &   16950-20340                                            &  8.2$\pm$0.1       & 0.7$\pm$0.2\\
                  &    S3   &   25990-30510; 31640-32770                       &  7.6$\pm$0.2       &1.1$\pm$0.1 \\
                  &    S4   &   24860-25990; 30510-31640; 32770-35030& 7.1$\pm$0.2        &1.2$\pm$0.2\\
 \\
 Obs 3       &    S1   & 565-2825;  8475-9605; 14125-15255          & 7.3$\pm$0.2              &1.2$\pm$0.2 \\
                  &   S2   &2825-6215; 7345-8475; 9605-11865; 12995-14125&   6.8$\pm$0.2  & 1.1$\pm$0.1\\
                  &   S3   & 6215-7345; 11865-12995; 15255-16385      &  6.1$\pm$0.5             &0.9$\pm$0.2 \\
\\
 Obs 4       & S1    & 0-3954           & 8.3$\pm$0.6 & 0.8$\pm$0.1 \\
                  & S2    & 3954-7344     & 6.7$\pm$0.2 & 1.5$\pm$0.1  \\
                  & S3    & 7344-11579   & 6.9$\pm$0.2 & 2.3$\pm$0.1  \\
                  & S4     & 37310-40800  & 9.4$\pm$0.3 & 1.3$\pm$0.1  \\
\hline
\end{tabular}
\medskip
\\
\label{segs}
\end{table*}

\subsection{Spectral data}
For the XMM-Newton observations we applied {\sc flag}=0 to exclude the events at the edge of the CCD and the ones close to a bad pixel to extract the spectra. We produced the response matrices using the task {\tt rmfgen} and the ancillary response files using the command {\tt arfgen}. For the latter we used the extended point spread function (PSF) model to calculate the encircled energy correction. Since the whole CCD was illuminated and contaminated by source photons, we extracted the background spectra from another XMM-Newton observation of the black hole candidate GX 339-4 in timing mode (ObsID 0085680601), based on similar sky coordinates and column density along the line of sight. For more details about the background extraction, please refer to \citet{hie11} and \citet{sanna13}. We rebinned the source spectra so that we have a minimum of 25 counts per bin, and oversampled the energy resolution of the PN detector by a factor of 3.

We used the {\sc heasoft} tools version 6.14 for the RXTE data reduction following the recipes in the RXTE cook book. We applied the tool {\tt saextrct} to generate the PCA spectra from Standard-2 data using only events from the best-calibrated PCU2 detector. We generated the background using the tool {\tt pcabackest} and the response files using {\tt pcarsp}, respectively. We extracted the HEXTE spectra for cluster-B events only using the script {\tt hxtlcurv}, and produced the response files using the command {\tt hxtrsp}. 

For Obs 1-3, we fitted the spectrum of each segment from EPIC-PN, PCA and HEXTE data simultaneously. For Obs 4, since the RXTE observation covered only segment 1 to 3, we combined these three RXTE segments to generate RXTE spectra for segment 4. Besides, we noticed that there was a moderately high background flare in the XMM-Newton Obs 3, so we applied flare filtering to produce spectra of the total observation. We did not apply flare filtering to spectra of the segments in this observation, otherwise we would have excluded the segment 3 since most exposure time of this segment was covered by high background flare. Alternatively, we only selected PN spectra below 5 keV for spectral analysis to remove the possible influence from the flare. 
 
We first fitted the PN spectra in the $0.8-10$ keV, and the PCA and HEXTE spectra in the range of $10-25$ and $20-100$ keV, respectively. We found that there were very few source photons in the HEXTE spectra above 30 keV in Obs 3. The parameters in the fits with HEXTE spectra up to 30 keV were exactly the same as in the fits with HEXTE spectra up to 100 keV, and the corresponding reduced chi-square values were much smaller. Therefore we decided to use the HEXTE spectra below 30 keV to fit the segments in Obs 3.

For the reduction of all X-ray bursts detected by the PCA, we produced a spectrum every 0.25 s during the whole duration of each X-ray burst.
We generated the instrument response matrix for each spectrum with the standard {\sc ftools}
routine {\sc pcarsp}, and we corrected each spectrum for dead time using the
methods supplied by the RXTE team. For each burst we extracted the spectrum of the
persistent emission just before or after the burst to use as background in our fits.

\section{Results}
\subsection{Timing results}
In Figure \ref{ps} we show the average power spectra of the XMM-Newton observations. We detected a strong QPO at about 8$-$10 mHz and a second harmonic component in each observation. In Figure \ref{dps} it is apparent that the mHz QPO and its second harmonic component disappeared right before an X-ray burst in all observations. The QPO in Obs 1 reappeared together with its second, and possible third, harmonics $\sim$ 21 ks later after the burst, and in Obs 4 the QPO reappeared $\sim$25 ks after the previous burst. In Obs 1 and 3 the QPO appeared from the beginning of the observations, while in Obs 2 the QPO appeared 16.4 ks after the start of the observation, and then lasted for $\sim$18.6 ks until it became undetectable before an X-ray burst. 

In Figure \ref{fre} we show the frequency of the QPO in the intervals of each XMM-Newton observation. In Obs 1, the frequency decreased from $\sim$9.9 mHz to $\sim$ 8.5 mHz before the burst, it reappeared at a frequency of $\sim$9.6 mHz, then it increased to $\sim$10.4 mHz and returned back to 9.6 mHz at the end of the observation. The frequency in Obs 2 started at $\sim$8 mHz, it increased to $\sim$9 mHz, and finally it decreased to below $\sim$7.6 mHz until an X-ray burst happened. Interestingly, the frequency in Obs 3 seemed to move randomly around $\sim$6.8 mHz before the burst. In Obs 4 the frequency decreased from $\sim$ 8.8 mHz to $\sim$ 6.5 mHz, then increased to $\sim$ 7 mHz before the burst, it reappeared at $\sim$9.7 mHz after the burst, and then decreased to $\sim$9.1 mHz at the end of the observation. The average frequency and rms amplitude of the QPO in each segment are shown in Table \ref{segs}; we found no clear correlation between the frequency and the rms amplitude.

\begin{figure}
\center
\includegraphics[height=0.65\textwidth,angle=-90]{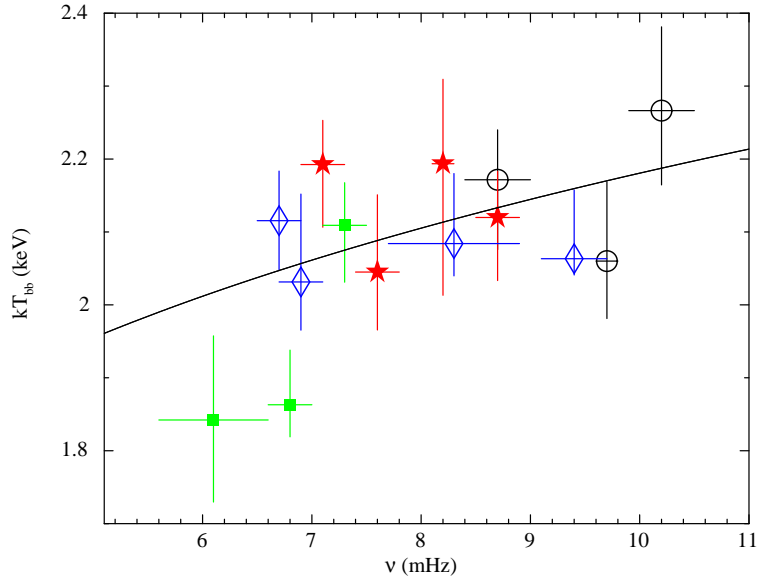}
\caption{
Average frequency of the mHz QPO vs. the temperature of the blackbody component in 4U 1636$-$53. The results for Obs 1-4 are marked with black circles, red stars, green squares and blue diamonds, respectively. Here we plot the standard deviation of the average frequency in each segment as the error bar of the frequency. The error bar of the temperature is at 68\% confidence level. The black curve in the plot corresponds to the best-fitting power-law model to the data when only the error bar of the temperature is used during the fitting process.
}
\label{cor}
\end{figure}

\subsection{Spectral results}
We fitted the spectra using XSPEC 12.8.1 \citep{arnaud96} with a model consisting of two thermal components and a Comptonised component \citep[e.g.,][]{sanna13,lyu14a}. We used the {\sc bbody} component to fit the thermal emission from the neutron star surface and its boundary layer. We described the thermal emission from the accretion disc by the {\sc diskbb} component \citep{mitsuda84,makis86}. For the Comptonised component we selected the model {\sc nthcomp} \citep{zdzi96,zyck99} to describe the non-thermal emission from the inverse Compton scattering process in the corona, with the seed photons coming from the disc \citep[e.g.,][]{sanna13,lyu14b,lyu14a}. We used the component {\sc phabs} to account for the interstellar absorption along the line of sight, selecting the solar abundance table of \citet{wilms00} and the cross section table of \citet{verner96}. We also used the component {\sc gauss} to account for the iron emission line around 6$-$7 keV in this source \citep[e.g.,][]{pandel08,Ng10,sanna13,lyu14a}. Additionally, we added a multiplicative {\tt constant} to the model to account for possible calibration differences between the instruments. Finally, a 0.5\% systematic error was added to the model.

For each observation, we assumed the column density, N$_{H}$, in {\sc phabs} in each segment to be the same and fixed it to be the one in the full observation. We found that the electron temperature, $kT_{e}$ in {\sc nthcomp}, was not well constrained due to the limited exposure time of each segment. To overcome this issue we fixed $kT_{e}$ in the segments to the value derived from the fits of the corresponding total observation. We fixed the normalisation of the {\sc diskbb} component in the segments of Obs 2 to the value in the total observation since this parameter in the segments could not be well constrained and pegged at its lower boundary. We did not use the {\sc gauss} component to fit the segments in Obs 3 since the PN spectra above 5 keV was removed in order to reduce background flare. The final fitting results of the total observations and the segments are shown in Table \ref{totfits} and \ref{fits}.

We analysed the spectra of all X-ray bursts in 4U 1636$-$53 using {\sc xspec} version 12.8.0 \citep{arnaud96}, restricting the
spectral fits to the energy range $3.0-20.0$ keV. We fitted the time-resolved spectra using
a single-temperature blackbody model ({\sc bbodyrad} in {\sc xspec}) times a component that
accounts for the interstellar absorption towards the source and fixing the hydrogen column density,
$N_{\rm H}$, to $0.36\times10^{22}$ cm$^{-2}$ \citep{sanna13}. The spectral model provides two
free parameters: the blackbody colour temperature ($T_{\rm bb}$) and the normalization, which is proportional
to the square of the blackbody radius of the emitting surface, and allows us to estimate
the bolometric flux during the burst. We then measured the burst fluence $E_{\rm b}$ by integrating
the measured fluxes over the burst interval. We used the characteristic timescale
$\tau =E_{\rm b}/ f_{peak}$, where $f_{peak}$ is the peak flux of the burst, to measure the duration of the bursts.

In Figure \ref{cor} we show the frequency of the mHz QPO vs. the temperature of the blackbody component in the four observations. The frequency covers from $\sim$6 mHz to $\sim$10 mHz, and the temperature, $kT_{bb}$, ranges between $\sim$1.9 keV and $\sim$2.3 keV. We first fitted these data with a constant model and got a reduced chi-square of 1.71 ($\chi^2/dof$=22.23/13). We also fitted these data with a power-law model, and got a power-law index of 0.16$\pm 0.08$ (1-$\sigma$ error) with a reduced chi-square of 1.50 ($\chi^2/dof$=18.09/12; for both fits we took into account only the errors in the temperature). The F-test probability for these two fits is 0.88, indicating that a power-law fit is not significant better than a constant. Since Obs 3 was partly affected by solar flares (see \S 2.2), we also fitted the data with a power-law model excluding the points coming from this observation. In this case the power-law index is $0.01 \pm 0.09$.

\begin{figure}
\center
\includegraphics[height=0.7\textwidth,angle=-90]{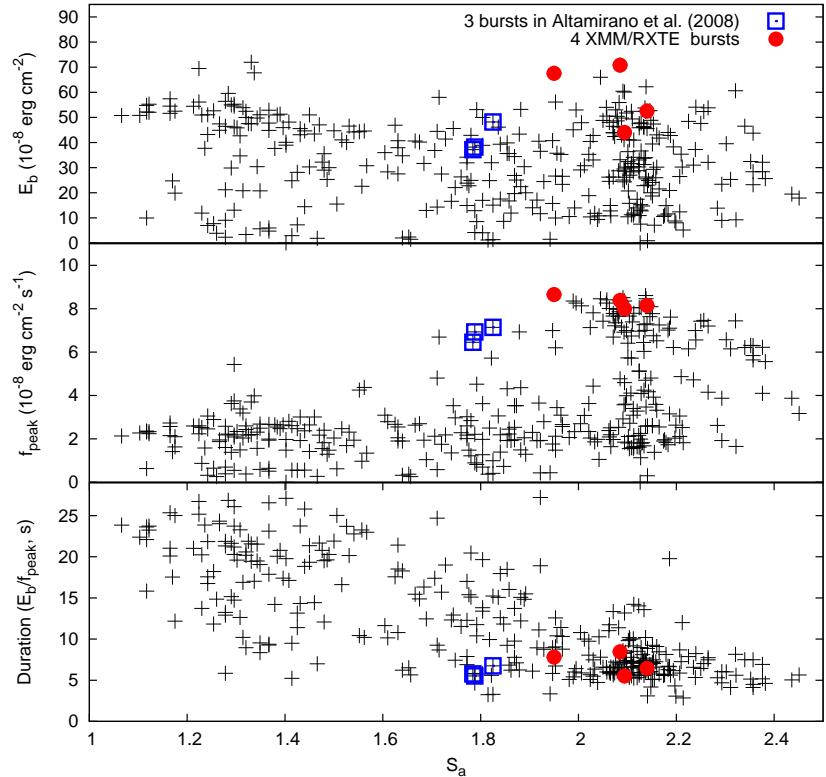}
\caption{Fluence (upper panel), peak flux (middle panel) and duration (bottom panel) of all X-ray bursts in 4U 1636$-$53 detected with the RXTE satellite as a function of S$_{\rm a}$ (position of the source in the colour-colour diagram; see $\S$2 for more details). The black crosses represent all 338 type-I X-ray bursts, while the blue squares and the red circles represent the seven bursts associated with mHz QPOs in \citet{diego08} and this work.
}
\label{burst}
\end{figure}

In Figure \ref{burst}, we show the properties of all X-ray bursts in 4U 1636$-$53 detected by the RXTE satellite as a function of the parameter S$_{\rm a}$. Among the $\sim$340 type I X-ray bursts detected with RXTE from this source, seven were confirmed to be associated with mHz QPOs: the three bursts reported in \citet{diego08} plus the four bursts in the XMM-Newton/RXTE observations in this work. The S$_{\rm a}$ of those 7 bursts range from 1.8 to 2.2, indicating that those bursts appeared around the transitional spectral state. From the upper panel of the figure it is apparent that the fluence of those bursts is relatively high, above $40\times 10^{-8}$ erg cm$^{-2}$. The middle panel of the figure shows that the peak flux of those 7 bursts is above $6\times 10^{-8}$ erg cm$^{-2}$ s$^{-1}$, which is much higher than the average peak flux of all bursts in this source. In addition, as shown in the bottom panel, those 7 bursts have relatively short durations.
 
\begin{table}
\caption{Fitting results of the observations of 4U 1636$-$53. Here we give the total unabsorbed flux from 0.5-100 keV in unit of erg cm$^{-2}$ s$^{-1}$.}
\begin{tabular}{|c|c|c|c|c|c|}
\hline
Model Comp & Parameter &  Obs1 & Obs2 & Obs3 & Obs4   \\
\hline                                                                                                                                                 
{\sc phabs}  &$N_{\rm H}$(10$^{22}$cm$^{-2}$)&  0.29$\pm 0.03      $&  0.32$\pm 0.02        $&  0.37$\pm 0.004        $&   0.38$\pm 0.004       $ \\
{\sc bbody}  &$kT_{\rm BB}$(keV)             &  2.25$\pm 0.15      $&  2.03$\pm 0.19        $&  2.01$\pm 0.02         $&   1.96$\pm 0.03        $ \\
             &Nor       (10$^{-3}$)          &  4.19$\pm 0.36      $&  3.32$\pm 0.91        $&  5.20$\pm 0.13         $&   5.28$_{-0.47}^{+0.15}$ \\
{\sc diskbb} & $kT_{dbb} (keV)$              &  0.66$\pm 0.05      $&  0.68$\pm 0.04        $&  0.64$\pm 0.02         $&   0.62$\pm 0.03        $ \\
             &Nor                            &    69$_{-55}^{+33}  $&    52$_{-32}^{+98}    $&    87$_{-3}^{+5}       $&     50$_{-31}^{+3}     $ \\
{\sc nthcomp}& $\Gamma$                      &  2.52$\pm 0.16      $&  2.31$_{-0.16}^{+0.28}$&  2.38$_{-0.08}^{+0.01} $&   2.49$_{-0.02}^{+0.2} $ \\
             &kT$_{\rm e}$(keV)              &  9.8$_{-2.9}^{+4.9} $&  3.7$_{-0.4}^{+0.9}   $&   4.2$\pm 0.1          $&    4.7$_{-0.4}^{+2.6}  $ \\
             & Nor                           &  0.31$\pm 0.09      $&  0.40$_{-0.19}^{+0.06}$&  0.42$\pm 0.005        $&   0.42$\pm 0.03        $ \\
{\sc gauss}  &$E_{\rm line} (keV)$           &  6.4 $_{-0}^{+0.2}   $&  6.55$\pm 0.15        $&  6.75$\pm 0.12         $&   6.80$\pm 0.06        $ \\
             &$\sigma (keV)$                 &  1.6$_{-0.4}^{+0.1} $&  1.5$\pm 0.2          $&   1.3$_{-0.2}^{+0.1}   $&   1.21$\pm 0.07        $ \\
             &Nor     (10$^{-3}$)            &  5.5$_{-2.2}^{+1.3} $&  6.2$_{-1.8}^{+2.4}   $&   4.0$_{-0.7}^{+0.2}   $&    4.1$\pm 0.6         $ \\
             &  $\chi^2/dof$    &  202/208       &  220/209          &   233/208          &   242/202           \\
             & Total flux (10$^{-10}$)   &  25.8$\pm 0.3$       &    31.6$\pm 0.2$  &   33.3$_{-0.7}^{+0.03}$  & 29.7$_{-0.6}^{+0.1}$           \\
            
\hline
\end{tabular}
\medskip
\\
\label{totfits}
\end{table}

\begin{table}
\tiny
\caption{Fitting results of the segments in the observations of 4U 1636$-$53. N$_{H}$ and $kT_{e}$ are fixed to the values in the corresponding total observations: 0.29$\times 10^{22}$cm$^{-2}$ and 9.8 keV for Obs 1, 0.32$\times 10^{22}$cm$^{-2}$ and 3.7 keV for Obs 2, 0.37$\times 10^{22}$cm$^{-2}$ and 4.2 keV for Obs 3 and 0.38$\times 10^{22}$cm$^{-2}$ and 4.7 keV for Obs 4, respectively. We also fix the normalisation of the {\sc diskbb} component in the individual segments of Obs 2 to the value in the total observation since this parameter in the segments can not be constrained in the fits. For Obs 3, we do not use the {\sc gauss} component at $\sim$6-7 keV since the PN spectra above 5 keV was removed to reduce background flare effects.}
\begin{tabular}{|c|c|c|c|c|c|c|c|c|c|c|}
\toprule
   &  \multicolumn{2}{c}{{\sc bbody}} & \multicolumn{2}{c}{{\sc diskbb}} &\multicolumn{2}{c}{{\sc nthcomp}} & \multicolumn{3}{c}{{\sc gauss}}\\

Obs.Seg  &  $kT (keV)$          &   Nor (10$^{-3}$)        &    $kT_{dbb} (keV)$     &   Nor     &   $\Gamma$            & Nor          &   $E_{\rm line} (keV)$     &   $\sigma (keV)$ &   Nor (10$^{-3}$)  &  $\chi^2/dof$\\
\hline
O1.S1   &2.27$\pm 0.18          $&4.3$\pm 0.7          $&0.69$\pm 0.03          $&69$\pm 36                 $&2.52$\pm 0.16          $& 0.30$\pm 0.07          $&6.65$\pm 0.22           $&1.0$_{-0.3}^{+0.5}    $&2.9$_{-0.9}^{+2.8}    $& 218/209 \\
O1.S2   &2.06$\pm 0.16          $&4.8$\pm 1.1          $&0.71$\pm 0.03         $&110$_{-33}^{+22}     $&2.39$\pm 0.12         $&0.23$\pm 0.06         $&6.44$_{-0.04}^{+0.24} $&1.2$_{-0.3}^{+0.6}   $&2.7$_{-1.3}^{+4.3}    $&   228/209 \\
O1.S3   &2.17$\pm 0.14          $&4.5$\pm 0.6          $&0.71$\pm 0.03         $&113$\pm 18               $&2.39$\pm 0.12          $&0.23$\pm 0.05        $&6.43$_{-0.03}^{+0.29}  $&1.3$\pm 0.4              $&3.6$_{-1.7}^{+2.9}    $& 183/208 \\
\hline
O2.S1   &2.12$\pm 0.12         $&5.3$\pm 1.1          $&0.74$\pm 0.03          $&  52                               &2.42$\pm 0.08           $&0.38$\pm 0.01        $&6.57$_{-0.17}^{+0.31}   $&1.2$\pm 0.4              $&4.5$_{-2.1}^{+3.5}   $& 194/209  \\
O2.S2   &2.19$\pm 0.25         $&2.4$\pm 1.0          $&0.69$\pm 0.03          $&  52                               & 2.34$\pm 0.06          $&0.40$\pm 0.01        $&6.4$_{-0}^{+0.29}            $&1.8$_{-0.3}^{+0.2}   $&11.2$\pm 3.1           $&  222/210 \\
O2.S3   &2.04$\pm 0.15         $&4.7$\pm 1.0          $&0.74$\pm 0.04          $&    52                             & 2.39$\pm 0.07          $&0.39$\pm 0.02        $&6.7$_{-0.3}^{+0.27}        $&1.4$\pm 0.5             $&4.9$_{-2.7}^{+5.1}    $&  240/210  \\
O2.S4   &2.19$\pm 0.12         $&5.1$\pm 1.0          $&0.76$\pm 0.03            $&   52                             &  2.44$\pm 0.08         $&0.38$\pm 0.01         $&6.51$_{-0.11}^{+0.29}   $&1.3$\pm 0.3             $&6.4$\pm 2.9               $& 223/210  \\
\hline
O3.S1  &2.11$\pm 0.12               $&7.5$\pm 1.4       $&0.65$\pm 0.03         $&67$_{-66}^{+46}       $&2.51$\pm 0.10         $&0.44$\pm 0.07         $&  - & - & - &  140/112   \\
O3.S2  &1.86$_{-0.07}^{+0.12} $& 5.2$\pm 0.7      $&0.65$\pm 0.03         $&134$\pm 37               $&2.31$\pm 0.05         $&0.36$\pm 0.06         $& - & -  &-  &  121/111   \\
O3.S3  &1.84$\pm 0.19              $& 5.2$\pm 0.9       $&0.66$\pm 0.04         $&158$_{-51}^{+32}     $& 2.25$\pm 0.08        $&0.32$\pm 0.06         $&  - & -& -  &  122/111   \\
\hline
O4.S1 &2.08$_{-0.07}^{+0.16}  $&4.0$\pm 0.7          $&0.63$\pm 0.03         $&40$_{-39}^{+92}     $&2.53$\pm 0.09               $&0.43$\pm 0.07         $&6.67$\pm 0.12                 $&1.3$\pm 0.2              $&6.4$_{-2.0}^{+4.0}   $& 234/202   \\
O4.S2 &2.12$\pm 0.11                $&5.1$\pm 0.5          $&0.65$\pm 0.03        $&99$_{-40}^{+24}      $&2.50$_{-0.06}^{+0.22} $&0.36$\pm 0.06         $&6.57$\pm 0.17                 $&1.1$_{-0.1}^{+0.2}   $&4.0$_{-1.3}^{+2.3}   $& 210/202   \\
O4.S3 &2.03$\pm 0.16                $&5.2$\pm 1.0          $&0.67$\pm 0.04         $&118$\pm 56             $&2.47$\pm 0.16               $&0.32$\pm 0.08         $&6.51$_{-0.11}^{+0.37}   $&1.5$\pm 0.3               $&4.7$\pm 2.6              $& 191/208   \\
O4.S4 &2.06$_{-0.04}^{+0.16}  $&5.4$\pm 0.7          $&0.65$\pm 0.02          $&57$_{-6}^{+58}       $&2.50$\pm 0.05               $&0.40$\pm 0.08         $&6.74$\pm 0.18                 $&1.1$\pm 0.2              $&3.7$_{-1.5}^{+2.5}   $& 216/202   \\
\bottomrule
\end{tabular}
\medskip
\\
\label{fits}
\end{table}

\section{Discussion}
We report three new detections of the mHz QPO in the neutron star LMXB 4U 1636--53 with XMM-Newton. The QPO in all observations in this work disappeared before an type I X-ray burst. In the first and the fourth observation, the QPO reappeared at the end of each observation after burst, showing a reappearance time of $\sim$ 21 ks and $\sim$25 ks, respectively. The QPO in the second observation appeared $\sim$16 ks after the beginning of the observation, and then lasted for $\sim$18.6 ks until it disappeared right before a X-ray burst. We found no significant correlation between the frequency of the mHz QPO and the temperature of the blackbody component in the XMM-Newton/RXTE observations. Besides, we also found that there was a connection between the mHz QPO and type I X-ray burst properties.

The mHz QPOs in the four observations in the transitional spectral state of 4U 1636--53 all disappeared right before X-ray bursts, consistent with the conclusion in \citet{diego08} that in the transitional state the mHz QPOs disappear only when an X-ray burst happens. However, the frequency behaviour before the burst in Obs 3 was different from the frequency drift observed in other observations in the transitional state \citep{diego08,lyu14b}.

\citet{diego08} reported a time scale of $>$15 ks for mHz QPO to reappear after the burst when the source was in the transitional state, and a reappearance time of $\sim$ 6 ks when the source was in the soft state. Considering the two complete QPO reappearance times in this work ($\sim$ 21 ks in Obs 1 and $\sim$ 25 ks in Obs 4), it appears that the reappearance time of mHz QPOs in the transitional state is longer than in the soft state. If that is the case, this may indicate that either the X-ray bursts burn more fuel or the fuel accumulation process after a burst is slower in the transitional state than in the soft state.

The frequency of the mHz QPO in Obs 2 first increases and then decreases for $\sim$15 ks until a burst happens. According to the model of \citet{keek09}, this frequency drift is determined by the cooling process of the burning layer, so the frequency drift time of 15 ks corresponds to the cooling timescale of the burning layer. Assuming a burning process with standard reaction rates from \citet{keek14}, we found that the thermal cooling timescale of 15 ks corresponds to a column depth of $2\times 10^{9}$ g cm$^{-2}$, which is an order of magnitude higher than where the stable H/He burning takes place, therefore, it is likely that the layer is heated by the X-ray burst before it cools down to trigger the frequency drift of mHz QPOs. 

According to \citet{heger07} (see their eq. 11), and assuming that the thickness of the fuel layer remains more or less constant during the mHz QPOs cycle, the neutron-star temperature (that we identify with the blackbody temperature in our fits) should be proportional to $(\nu/\dot m)^{-2}$, where $\nu$ is the frequency of the mHz QPO, and $\dot m$ is the mass accretion rate onto the neutron star. Assuming that the total unabsorbed flux ($F$) is proportional to mass accretion rate, we fitted $kT_{BB}$ vs. $\nu/F$ with a power law relation and got a power law index of 0.11$\pm 0.05$ (1-$\sigma$ error) with a reduced chi-square of 1.4 ($\chi^2/dof$=17.33/12). This number is significantly different from the index of $-2$ predicted by \citet{heger07}.

The simulations of \citet{keek09} suggest that there should be a correlation between the frequency of the mHz QPO and the temperature of the blackbody component: As the burning layer gradually cooles down, the frequency of the oscillations decreases by tens of percents until an X-ray burst happens. To investigate how our observational constraints on the power-law
index compare to multi-zone simulations, we considered the hydrogen-accreting
models presented by \citet{keek14}. We determine the power-law
index $\beta$ for the different series of models where marginally
stable burning was present by using the oscillation period reported
by \citet{keek14} and the time-averaged effective surface
temperature of the models. For most models $\beta\simeq0.13$, whereas
the simulations with a decreased rate of the $^{15}\mathrm{O}\left(\alpha,\gamma\right)\mathrm{^{19}Ne}$
reaction yield $\beta=0.22$. A study of pure-helium burning models
yields $\beta\simeq0.04$ (\citealt{keek09} for $\dot{M}=0.3\,\dot{M}_{\mathrm{Edd}}$,
without rotationally induced mixing). Our observational constraint
on $\beta$ is consistent with these values. Interestingly, the predicted
change in $\beta$ for the reduced $^{15}\mathrm{O}\left(\alpha,\gamma\right)\mathrm{^{19}Ne}$
rate is similar to the $1\sigma$ uncertainty in our measurement.
Future improved observations of mHz QPOs hold a promise to constrain
the uncertain nuclear reaction rates of CNO breakout reactions \citep[e.g.,][]{Davids2011}
that have been shown to have a great effect on Type I bursts as well
as other burning processes \citep{Fisker2006,Cooper2006,Fisker2007,Parikh2008,Davids2011}.
Besides nuclear reaction rates and accretion composition, the stability
transition depends on a series of parameters that have not been exhaustively
studied, such as the base luminosity and rotationally induced mixing
\citep{keek09}, as well as the effective surface gravity \citep{heger07}.

\citet{Galloway08} showed that the long bursts with slow rise and decay were likely mixed H/He rich bursts, while fast bursts with shorter rise time were likely to burn primarily Helium. \citet{sugimoto84} and \citet{galloway06} investigated the distribution of peak fluxes of bursts in 4U 1636$-$53; these results show that some faint photospheric radius expansion (PRE) bursts have relatively low peak fluxes. According to their interpretation, during the expansion of the photosphere, a H-rich outer layer is ejected, exposing the underlying helium layer below. For the normal PRE bursts, the bursts reach the $L_{\rm Edd,He}$ after the ejection of the outer H-rich layer, while the faint PRE bursts have insufficient energy to drive off the H-rich layer, reaching only $L_{\rm Edd,H}$ instead of $L_{\rm Edd,He}$. In this work, all seven PRE bursts associated with mHz QPOs are short, bright and energetic, suggesting that they reach $L_{\rm Edd,He}$ and hence belong to He-rich bursts. Therefore, it is very likely that there is potential connection between the mHz QPOs and the He-rich PRE bursts when the source is in the transitional state.

\section*{Acknowledgments}
This research has made use of data obtained from the High Energy Astrophysics Science Archive Research Center (HEASARC), provided by NASA's Goddard Space Flight Center. This research made use of NASA's Astrophysics Data System. LM is supported by China Scholarship Council (CSC), grant number 201208440011. LK acknowledges support from NASA ADAP grant NNX13AI47G.

\clearpage
\bibliographystyle{mn}
\bibliography{biblio}

\label{lastpage}

\end{document}